# Effect of slip boundary conditions on interfacial stability of two-layer viscous fluids under shear

Stanislav Patlazhan

*Semenov Institute of Chemical Physics of Russian Academy of Sciences, 4, Kosygin Street, 119991 Moscow, Russia*

The traditional approach in the study of hydrodynamic stability of stratified fluids includes the stick boundary conditions between layers. However, this rule may be violated in polymer systems and as a consequence various instabilities may arise. The main objective of this paper is to analyze theoretically the influence of slip boundary conditions on the hydrodynamic stability of the interface between two immiscible viscous layers subjected to simple shear flow. It is found that the growth rate of long-wave disturbances is fairly sensitive to the slip at the interface between layers as well as at the external boundary. These phenomena are shown to give different contributions to the stability of shear flow depending on viscosity, thickness, and density ratios of the layers. Particularly, the interfacial slip can increase the perturbation growth rate and lead to unstable flow. An important consequence of this effect is the violation of stability for sheared layers with equal viscosities and densities in a broad range of thickness ratios. The conditions of long-wave propagation are analyzed as well, showing a strong dependence on the type of boundary conditions.

E-mail: sapat@yandex.ru

# I. INTRODUCTION

Several observations were recognizing slip of flowing polymer melts near solid walls[1-7] as well as at polymer-polymer interfaces.[8] Slip was also mentioned in water confined between hydrophobic surfaces.[9-11] Discontinuity of the velocity profile at the slippery interfaces results in abnormally low viscosity observed both for water and immiscible polymer blends.[12-14] The physical origin of slip depends on properties of both fluids and contacting surfaces. For instance, in the case of a polymer melt, wall slip is basically a function of the contact energy and of the molecular weight of the polymer. The first parameter controls the adhesion strength while the chain length controls the entanglements with macromolecules in the bulk. If the characteristic adhesion time is much smaller than the disentanglement time, slip may appear only due to the failure of adhesion.[15] In the case of confined water, adhesive slip may arise as a result of the competition between hydrogen bonding in the bulk and weak interaction with the hydrophobic surface. On the contrary, if interactions with the solid surface become significant, the slip boundary conditions may occur only for polymer fluids. Brochard and de Gennes[16] pointed out that disentanglement of adhered chains with the remaining polymer melt could result in cohesive slip.

Slip at the interface between immiscible polymer melts is caused by a sharp decrease in the interfacial viscosity. Its origin is conditioned by different mobilities of polymer chains at the interface and in the bulk. Indeed, assuming that contacting polymers obey Rouse dynamics, it was argued that the interfacial viscosity is proportional to the average length of polymer loops penetrating into the opposite phase, which are much less than the total chain length.[17-19]

The simplest slip boundary condition at the flat interface between two fluids (see Fig. 1) is given by Navier's ratio

$$\hat{u}_1^{(1)}(0) - \hat{u}_1^{(2)}(0) = \frac{\hat{l}_1}{\eta_1}\hat{\sigma}_{12}^{(1)}(0) \qquad (1)$$

describing the discontinuity of fluid velocity in terms of the applied shear stress $\hat{\sigma}_{12}^{(1)}(0)$. We denote $\hat{u}_1^{(\alpha)}$ as the horizontal projection of the velocity of the α-layer; $\hat{l}_1$ is the interfacial slippage length corresponding to the extrapolation of the upper fluid velocity to zero. The higher the slippage length, the larger the discontinuity of the velocity profiles; $\hat{l}_1$ is equal to zero under the non-slip boundary condition. The slippage length of a pair of immiscible polymers depends on the value of the Flory-Huggins parameter and can amount to 45 μm.[8] But in the case of a fluid-solid interface the slippage length may reach hundreds of microns in the case of high molecular-weight polymers.[5,15]

Apart from shear thinning, slip was shown to influence notably the stability of capillary flow and the shape of a polymer extrudate.[6,20-23] Particularly, depending on the flow rate, the wall slip of linear polymer melts at the capillary die may lead to a spurt or oscillatory flow. These flow instabilities can generate roughness at the free surface (known as sharkskin) or even gross distortions of the extrudate. Though much attention was devoted to capillary instabilities, the influence of slip on the stability of the interface between sheared immiscible fluids has not yet been studied. Among a great deal of papers analyzing the interfacial stability of different stratified fluids[24-29] attention was paid solely to the non-slip boundary conditions. Meanwhile there are several indications of correlation of hydrodynamic interfacial stability with the type of boundary conditions. This is suggested, for instance, by essentially different morphologies observed in reactive and nonreactive polymer blends after extrusion. It may be considered as a manifestation of the strong influence of modified boundary conditions on the nascent structure.

The main goal of this work is to show how the slip boundary conditions would alter the interfacial stability of a two-layer viscous fluid system subjected to low Reynolds number simple shear flow. For this purpose we generalize the long-wave perturbation theory, which provides stability diagrams showing the evolution of interfacial disturbances in terms of slippage lengths and system parameters such as ratios of viscosities, thicknesses, and densities of the layers. The important outcome of this study is that slip at the interface between viscous fluids promotes interfacial instability while slip at a solid wall may reduce it down to the neutral stability.

The paper is organized as follows. The governing equations and boundary conditions are considered in section II. In the third section the regimes of shear flow with slip boundary conditions are discussed. They include the primary flow and the flow perturbed by the long-wave disturbances. The fourth section is devoted to the perturbation analysis of the eigenvalue problem for propagation of interfacial waves. The wave speed and the amplitude growth rate of the disturbances are derived there. The analysis of propagation of disturbances and stability diagrams are discussed in the last part of the paper.

## II. GOVERNING EQUATIONS AND BOUNDARY CONDITIONS

Let us consider two superposed layers of incompatible and incompressible Newtonian fluids confined between parallel solid plates. Each layer is characterized by its viscosity $\eta_\alpha$, density $\rho_\alpha$, and thickness $d_\alpha$ with index $\alpha$ = 1, 2 corresponding to the upper and the lower layer respectively (see Fig. 1). The lower plate is kept fixed while the upper one moves along the $\hat{x}_1$-axis with a constant velocity $V$ generating simple shear flow. We are going to examine the two-dimensional flow described by the horizontal $\hat{u}_1$ and the vertical $\hat{u}_2$ projections of the fluid velocity. The origin of the reference frame is located at the undisturbed interface between the layers. Hence the upper and the lower walls are situated at $\hat{x}_2 = d_1$ and $\hat{x}_2 = -d_2$ respectively.

The governing Navier-Stokes equations for the upper and the lower viscous layers can be presented in the dimensionless form:

$$\frac{Du_i^{(1)}}{Dt} = -\frac{\partial p}{\partial x_i} + R_1^{-1}\Delta u_i^{(1)}$$
$$\frac{Du_i^{(2)}}{Dt} = -r^{-1}\frac{\partial p}{\partial x_i} + R_2^{-1}\Delta u_i^{(2)} \quad (2)$$

with the dimensionless coordinates $x_i$, velocities $u_i^{(\alpha)}$, pressure $p$, and time $t$ defined by the following expressions

$$x_i = \hat{x}_i/d_1,\ u_i^{(\alpha)} = \hat{u}_i^{(\alpha)}/V,\ p = \hat{p}/(\rho_1 V^2),\ t = \hat{t}V/d_1. \quad (3)$$

Index $i$ = 1, 2 denotes the type of axis. The Reynolds numbers of the upper and lower layers are defined as $R_1 = \rho_1 d_1 V \eta_1^{-1}$ and $R_2 = rm^{-1}R_1$ where the viscosity and density ratios, $m = \eta_2/\eta_1$ and $r = \rho_2/\rho_1$, are introduced. The differential operators $\frac{D}{Dt} = \frac{\partial}{\partial t} + u_1\frac{\partial}{\partial x_1} + u_2\frac{\partial}{\partial x_2}$ and $\Delta = \frac{\partial^2}{\partial x_1^2} + \frac{\partial^2}{\partial x_2^2}$ correspond to the convective time derivative and 2D Laplacian respectively. The positions of the upper and lower plates correspond to $x_2 = 1$ and $x_2 = -n$, where $n = d_2/d_1$ is the thickness ratio.

The incompressibility condition for the fluids is described by the equations

$$\frac{\partial u_1^{(\alpha)}}{\partial x_1} + \frac{\partial u_2^{(\alpha)}}{\partial x_2} = 0 \qquad (\alpha = 1, 2) \qquad (4)$$

The velocity profile of the considered two-phase system depends significantly on boundary conditions at the outer walls and the interface between the fluid layers. Assuming the non-slip boundary condition at the upper plate, we obtain the following projections of the surface velocity

$$u_1^{(1)}(x_1, 1) = 1 \text{ and } u_2^{(1)}(x_1, 1) = 0 \qquad (5)$$

On the other hand, we suppose that slip takes place at the lower wall as well as at the interface between the layers. We will assume that the discontinuities of the tangent velocities at these interfaces are proportional to the applied stress according to Navier-type boundary conditions. At the lower wall we thus obtain:

$$u_1^{(2)}(x_1, -n) = l_2 \left.\frac{\partial u_1^{(2)}}{\partial x_2}\right|_{x_2 = -n} \qquad (6a)$$

while the vertical projection of the velocity is

$$u_2^{(2)}(x_1, -n) = 0. \qquad (6b)$$

Parameter $l_2 = \hat{l}_2 / d_1$ corresponds to the dimensionless slippage length of the second layer at the lower plate.

The boundary conditions at the interface between the layers must take into account the local curvature of the disturbed interface $x_2 = y(x_1, t)$. For this reason Eq. (1), corresponding to the undisturbed interface $x_2 = 0$, should be modified: instead of the Cartesian projections of velocities and stresses, the tangent and normal components to an arbitrary point of the disturbed interface should be considered. The following boundary conditions are then obtained

$$u_t^{(1)}(x_1, y) - u_t^{(2)}(x_1, y) = (l_1 / \eta_1) \sigma_{tn}^{(1)}(x_1, y), \qquad (7a)$$

$$u_n^{(1)}(x_1, y) = u_n^{(2)}(x_1, y) \qquad (7b)$$

where $t$ and $n$ denote the tangent and normal projections. The first equation represents the Navier slip boundary condition with dimensionless slippage length $l_1 = \hat{l}_1 / d_1$. For the undisturbed interface Eq. (7a) coincides with Eq. (1). Note that depending on the thickness $d_1$ of the upper layer the dimensionless slippage lengths $l_1$ and $l_2$ may take on wide-ranging values. Hence one may tentatively conclude that the slip should not affect the flow under the conditions $d_1 \gg \hat{l}_1$, $\hat{l}_2$. On the contrary, a significant variation of the interfacial stability is expected for sufficiently thin layers.

The stress boundary conditions involve (i) the continuity of the shear stress and (ii) the balance of the normal stresses with the Laplace pressure due to the interfacial tension $\Gamma$ at the disturbed interface, $x_2 = y(x_1, t)$:

$$\sigma_{tn}^{(1)}(x_1, y) = \sigma_{tn}^{(2)}(x_1, y), \tag{8a}$$

$$\tau_{nn}^{(1)}(x_1, y) - \tau_{nn}^{(2)}(x_1, y) = -\frac{\Gamma}{d_1} \frac{\partial^2 y}{\partial x_1^2}. \tag{8b}$$

We took here into account the fact that the full stress tensor of the $\alpha$-layer depends on the pressure $p_\alpha$, gravity $g$, and viscous stress $\sigma_{ik}^{(\alpha)}(x_1, y)$ according to $\tau_{ik}^{(\alpha)}(x_1, y) = -p_\alpha \delta_{ik} + g y d_1 V^{-2} \delta_{i2} \delta_{k2} + \sigma_{ik}^{(\alpha)}$. Particularly, its normal component is equal to $\tau_{nn}^{(\alpha)}(x_1, y) = -p_\alpha + g y d_1 V^{-2} \cos^2 \beta + \sigma_{nn}^{(\alpha)}$ where $\beta$ is the angle between the tangent and the $x_2$-axis at an arbitrary point of the interface.

## III. REGIMES OF FLOW

### A. The primary flow

The primary flow corresponds to the undisturbed interface. In this case the velocity profile is described by the horizontal component $u_1^{(\alpha)} = U^{(\alpha)}$ (see Fig. 1). The Navier-Stokes equations (2) indicate that in this case the pressure $p$ is constant and the velocity is the linear function

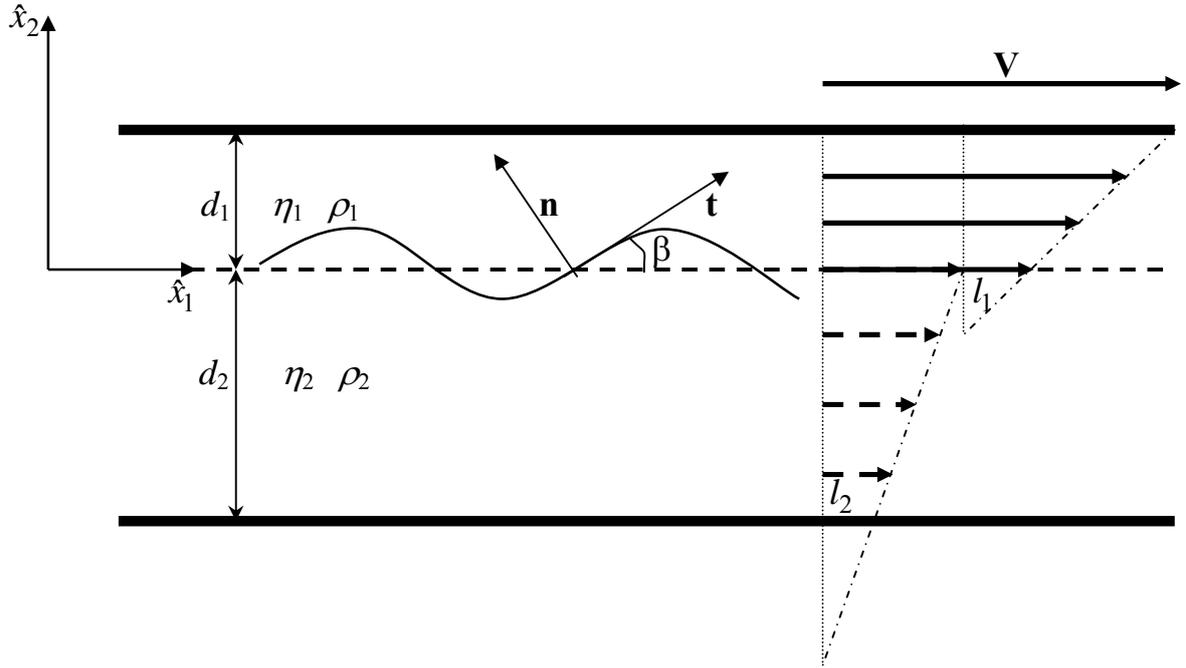

**Fig. 1.** The sketch of the shearing stratified system under the slip boundary conditions at the interface between the fluids and the lower wall. The upper plate moves with a constant velocity $V$. The slippage length $l_1$ and $l_2$ extrapolate relative boundary velocities to zero. The unit vectors **n** and **t** correspond to the normal and tangent at arbitrary point of the disturbed interface.

$$U^{(\alpha)}(x_2) = \dot{\gamma}_\alpha x_2 + V_\alpha \qquad (\alpha = 1, 2) \qquad (9)$$

where $\dot{\gamma}_\alpha$ is the shear rate and $V_\alpha$ the interface velocity of the $\alpha$-layer. In this case the boundary conditions (5), (6), (7) and (8) degenerate into the following expressions

$$U^{(1)}(1) = 1, \qquad (10)$$

$$U^{(2)}(-n) = l_2 \left.\frac{\partial U^{(2)}}{\partial x_2}\right|_{x_2 = -n}, \qquad (11)$$

$$U^{(1)}(0) - U^{(2)}(0) = l_1 \left.\frac{\partial U^{(1)}}{\partial x_2}\right|_{x_2 = 0}, \quad \left.\frac{\partial U^{(1)}}{\partial x_2}\right|_{x_2 = 0} = m \left.\frac{\partial U^{(2)}}{\partial x_2}\right|_{x_2 = 0}. \qquad (12)$$

Substituting Eq. (9) into Eqs. (10) - (12) and solving the obtained equations we find the following characteristic parameters of the primary flow:

$$\dot{\gamma}_1 = m\dot{\gamma}_2, \quad \dot{\gamma}_2 = (m + n + ml_1 + l_2)^{-1}, \tag{13}$$

$$V_1 = (n + ml_1 + l_2)\dot{\gamma}_2, \quad V_2 = (n + l_2)\dot{\gamma}_2 \tag{14}$$

It is seen that the shear rates in the layers decrease with slippage lengths and that the velocity discontinuity is proportional to $l_1$, i.e. $\Delta V = V_1 - V_2 = ml_1\dot{\gamma}_2$. For the non-slip boundary conditions, $l_1 = l_2 = 0$, we get the well-known classical results.

## B. The perturbed flow

The disturbance of the interface between the stratified fluids results in a perturbation of the fluid velocity in the layers. The total velocity is a sum of the primary and the perturbed velocities:

$$u_1^{(\alpha)}(x_1, x_2) = U^{(\alpha)}(x_2) + \tilde{u}_1^{(\alpha)}(x_1, x_2) \text{ and } u_2^{(\alpha)}(x_1, x_2) = \tilde{u}_2^{(\alpha)}(x_1, x_2) \tag{17}$$

A small periodical disturbance of the interface, $y(x_1, t) = y_0 \exp[ik(x_1 - ct)]$, creates a periodical perturbation of the velocity, $\tilde{u}_i^{(\alpha)}(x_1, x_2) = \tilde{u}_{0i}^{(\alpha)}(x_2)\exp[ik(x_1 - ct)]$ with dimensionless wavenumber $k = 2\pi d_1 / \lambda$. The long-wave limit $\lambda \gg d_1$ will be considered in this paper.

The regime of flow is generally defined by a complex number $c = c_0 + i\Delta c$. The real part of this value corresponds to the speed of wave propagation while the imaginary part characterizes the growth rate of the perturbation amplitude, which controls the interfacial stability. If $\Delta c$ is positive, the amplitude is growing up leading to unstable flow. Otherwise the disturbances are damped down.

The tangential and the normal velocities at the wavy interface $y(x_1, t)$ are given by the following relationships:

$$u_t^{(\alpha)}(x_1, x_2 = y) = u_1^{(\alpha)}(x_1, y)\cos\beta + u_2^{(\alpha)}(x_1, y)\sin\beta$$

$$\tag{18}$$

$$u_n^{(\alpha)}(x_1, x_2 = y) = -u_1^{(\alpha)}(x_1, y)\sin\beta + u_2^{(\alpha)}(x_1, y)\cos\beta$$

The angle $\beta$ between the tangent at the interface and the $x_1$-axis is equal to $\beta = \arctan(iky)$, which reduces to $\beta \cong iky$ in the long-wave limit. Taking that into account and neglecting the non-linear terms, we finally obtain:

$$u_t^{(\alpha)}(x_1,y) = U^{(\alpha)}(y) + \tilde{u}_t^{(\alpha)}(x_1,y), \qquad \tilde{u}_t^{(\alpha)}(x_1,y) = \dot{\gamma}_\alpha y(x_1,t) + \tilde{u}_1^{(\alpha)}(x_1,0) \qquad (19)$$

$$u_n^{(\alpha)}(x_1, x_2 = y) = \tilde{u}_n^{(\alpha)}(x_1,y), \qquad \tilde{u}_n^{(\alpha)}(x_1,y) = -iky(x_1,t)V_\alpha + \tilde{u}_2^{(\alpha)}(x_1,0) \qquad (20)$$

The same approach can be used for the derivation of the perturbations of shear and normal stresses at the interface:

$$\tilde{\sigma}_{tn}^{(\alpha)}(x_1,y) \cong \tilde{\sigma}_{12}^{(\alpha)}(x_1,0) \text{ and } \tilde{\sigma}_{nn}^{(\alpha)}(x_1,y) \cong -2ik\mu_\alpha \dot{\gamma}_\alpha y(x_1,t) + \tilde{\sigma}_{22}^{(\alpha)}(x_1,0) \qquad (21)$$

Using the incompressibility condition (4), introduce now the stream function $\psi_\alpha(x_1, x_2)$ of the $\alpha$-layer:

$$\tilde{u}_1^{(\alpha)} = \frac{\partial \psi_\alpha}{\partial x_2}, \quad \tilde{u}_2^{(\alpha)} = -\frac{\partial \psi_\alpha}{\partial x_1} \qquad (22)$$

In the case of periodical disturbances $\psi_\alpha(x_1,x_2) = \phi_\alpha(x_2)\exp[ik(x_1 - ct)]$. Taking into account the fact that the vertical projection of the interfacial velocity coincides with the perturbation rate, $\tilde{u}_2^{(\alpha)}(x_1,0) = \frac{dy}{dt} = -ik(c - V_\alpha)y(x_1)$, the amplitude of the stream function, $\phi_\alpha(0)$, at the interlaminar interface can be expressed as:

$$y_0 = \bar{c}_\alpha^{-1}\phi_\alpha(0), \qquad (23)$$

where $\bar{c}_\alpha = c - V_\alpha$ is the relative wave velocity of the $\alpha$-layer. Substituting this formula into Eqs. (20) and (21), one obtains the tangent and the normal components of the interface perturbation velocity expressed in terms of the stream function:

$$\tilde{u}_{0t}^{(\alpha)}(y) = \dot{\gamma}_\alpha \bar{c}_\alpha^{-1}\phi_\alpha(0) + \phi_\alpha'(0) \qquad (24)$$

$$\tilde{u}_{0n}^{(\alpha)}(y) = -ik\bar{c}_\alpha^{-1}c\phi_\alpha(0) \qquad (25)$$

The amplitudes of the shear and normal viscous stresses at the disturbed interface are then equal to

$$\tilde{\sigma}_{0tn}^{(\alpha)}(y) \cong \mu_\alpha \left\{ \phi_\alpha''(0) + k^2 \phi_\alpha(0) \right\} \tag{26}$$

$$\tilde{\sigma}_{0nn}^{(\alpha)}(y) \cong -2ik\mu_\alpha \left\{ \dot{\gamma}_\alpha \bar{c}_\alpha^{-1} \phi_\alpha(0) + \phi_\alpha'(0) \right\} \tag{27}$$

The Orr-Sommerfeld equation governing the behavior of the α-layer stream function is obtained by substitution of Eq. (22) into the Navier-Stokes equations (2):

$$\phi_\alpha^{iv} - 2k^2 \phi_\alpha'' + k^4 \phi_\alpha = -ikR_\alpha \left( c - U^{(\alpha)} \right)\left( \phi_\alpha'' - k^2 \phi_\alpha \right) \quad (\alpha = 1, 2). \tag{28}$$

The primes indicate derivatives of amplitude $\phi_\alpha(x_2)$ of the α-layer stream-function. The boundary value problem should be represented in this case in terms of stream functions. The values of the stream functions at the upper and lower walls follow from Eqs. (5) and (6):

$$\phi_1(1) = 0, \; \phi_1'(1) = 0 \tag{29}$$

$$\phi_2(-n) = 0, \; \phi_2'(-n) = l_2 \phi_2''(-n) \tag{30}$$

The boundary conditions at the interface between the layers are obtained through substitution of Eqs. (24) – (27) into Eqs. (7) and (8). This results in the following equations

$$-l_1 \phi_1''(0) + \phi_1'(0) + \left\{ (\dot{\gamma}_1 - \dot{\gamma}_2)\bar{c}_1^{-1} - l_1 k^2 \right\} \phi_1(0) = \phi_2'(0) \tag{31}$$

$$\frac{\phi_1(0)}{\bar{c}_1} = \frac{\phi_2(0)}{\bar{c}_2} \tag{32}$$

$$\phi_1''(0) + k^2 \phi_1(0) = m \left\{ \phi_2''(0) + k^2 \phi_2(0) \right\} \tag{33}$$

$$\begin{aligned} m\left(\phi_2'''(0) - 3k^2 \phi_2'(0)\right) - \left(\phi_1'''(0) - 3k^2 \phi_1'(0)\right) + \\ + ikR_1\{r(\bar{c}_2 \phi_2'(0) + \dot{\gamma}_2 \phi_2(0)) - (\bar{c}_1 \phi_1'(0) + \dot{\gamma}_1 \phi_1(0))\} = ikR_1\left((r-1)G + k^2 \Gamma_1\right)\bar{c}_1^{-1} \phi_2(0) \end{aligned} \tag{34}$$

where $G = g d_1 V^{-2}$ and $\Gamma_1 = \Gamma\left(d_1 \rho_1 V^2\right)^{-1}$. The term containing the interfacial tension $\Gamma$ in the right hand side of Eq. (34) is of order of $k^3$. By this reason it does not contribute to the long-wave problem. It is interesting to note that Eq. (32) accounts for a discontinuity of the stream functions at the interface between the layers. This is not the case of sticky boundary conditions.

Eqs. (29) – (34) constitute the full set of equations necessary to solve the problem of hydrodynamic stability of the interface between the viscous layers in the presence of slip and subjected to a small-amplitude long-wave perturbation.

## IV. PERTURBATION ANALYSIS

The stability of the stratified system under small Reynolds number shear flow will be treated by means of the perturbation theory. In this case the amplitudes of the stream functions and the wave speed can be expanded into series of the small parameter $kR_\alpha$:

$$\phi_\alpha = \phi_{0\alpha} + \phi_{1\alpha} + ... \text{ and } c = c_0 + i\Delta c + ... \tag{35}$$

The terms $\phi_{0\alpha}$ and $c_0$ represent the zero-order stream functions and the speed of the interfacial waves respectively, while $\phi_{1\alpha}$ and $\Delta c$ are the corresponding first-order in $kR_\alpha$ corrections of these values

### A. The zero-order approximation

We start with the infinite wavelength corresponding to zero value of the wavenumber $k$. In this limit the Orr-Sommerfeld equations (28) degenerate to

$$\phi_{0\alpha}^{iv} = 0 \tag{36}$$

leading to a cubical polynomial as the zero-order solution:

$$\phi_{0\alpha}(x_2) = A_\alpha + B_\alpha x_2 + C_\alpha x_2^2 + D_\alpha x_2^3 \tag{37}$$

In this case the boundary conditions (29) and (30) at the outer walls remain as before while Eqs. (31)-(34) are transformed into the following equations

$$\bar{c}_{02}\phi_{01}(0) = \bar{c}_{01}\phi_{02}(0) \tag{38}$$

$$\phi_{01}''(0) = m\phi_{02}''(0) \tag{39}$$

$$-l_1\phi_{01}''(0)+\phi_{01}'(0)+\frac{a_1-a_2}{\bar{c}_{01}}\phi_{01}(0)=\phi_{02}'(0) \tag{40}$$

$$\phi_{01}'''(0)=m\phi_{02}'''(0) \tag{41}$$

It is seen herefrom that the zero-order stream functions are independent of the density ratio. Substituting Eq. (37) into Eqs. (29), (30), (38) – (41) and solving the obtained set of equations, we arrive to the expression of the relative speed of the long-wave propagation

$$\bar{c}_{01}=c_0-V_1=\frac{m\{2n(m-1)((2+3n)l_2+n+n^2)-4m^2l_1^2-(2(2+3n)l_2+m+4n+3n^2)ml_1\}}{(m+n+ml_1+l_2)\{4(3n^2l_2+m+n^3)ml_1+4(m(3n+3n^2+1)+n^3)l_2+m^2+n^4+2mn(2+3n+2n^2)\}} \tag{42}$$

taking into account the slip boundary conditions both at the interface between the layers and at the lower wall The expressions for the coefficients of the zero-order stream functions (37) in terms of $m$, $n$, $l_1$ and $l_2$ are given in the Appendix.

In the case of the non-slip boundary conditions Eq. (42) reduces to the well-known expression:[24]

$$\bar{c}_0(l_1=l_2=0)=\frac{2mn^2(m-1)(1+n)}{(m+n)\{m^2+n^4+2mn(2+3n+2n^2)\}} \tag{43}$$

The consequences of the slip boundary conditions on the propagation of the perturbation waves will be discussed in the next section.

**B. The first-order approximation**

In this approximation only terms linear in wavenumber $k$ are kept. In this case the Orr-Sommerfeld equation (28) is written as

$$\phi_{1\alpha}^{iv}+ikR_\alpha(c_0-U^{(\alpha)})\phi_{0\alpha}''=0 \qquad (\alpha=1,2). \tag{44}$$

Taking Eq. (37) into account we obtain the first-order corrections to the $\alpha$-layer stream function:

$$\phi_{1\alpha}(x_2)=ikR_\alpha\left[\Delta A_\alpha+\Delta B_\alpha x_2+\Delta C_\alpha x_2^2+\Delta D_\alpha x_2^3+h_\alpha(x_2)\right] \tag{45}$$

with

$$h_\alpha(x_2) = \frac{1}{60}\dot{\gamma}_\alpha D_\alpha x_2^6 + \frac{1}{60}(\dot{\gamma}_\alpha C_\alpha - 3\bar{c}_{0\alpha} D_\alpha)x_2^5 - \frac{1}{12}\bar{c}_{0\alpha} C_\alpha x_2^4 \qquad (46)$$

These expressions coincide formally with those obtained by Yih (1967) except that the coefficients $C_\alpha$ and $D_\alpha$ in Eq. (46) depend now on the interfacial and wall slippage lengths, $l_1$ and $l_2$ (see Appendix).

The boundary conditions at the external walls for the first order corrections $\phi_{1\alpha}(x_2)$ follow from Eqs. (29) and (30):

$$\phi_{11}(1) = 0, \; \phi_{11}'(1) = 0 \qquad (47)$$

$$\phi_{12}(-n) = 0, \; \phi_{12}'(-n) = l_2 \phi_{12}''(-n) \qquad (48)$$

At the interface between the layers they are derived from Eqs. (31) – (34). Taking into account Eqs (38) – (41) we arrive to:

$$\bar{c}_{01}\bar{c}_{02}\phi_{11}(0) - i\Delta c(b_1 - b_2)\phi_{01}(0) = \bar{c}_{01}^2 \phi_{12}(0) \qquad (49)$$

$$\phi_{11}''(0) = m\phi_{12}''(0) \qquad (50)$$

$$-l_1\phi_{11}''(0) + \phi_{11}'(0) + \frac{\dot{\gamma}_1 - \dot{\gamma}_2}{\bar{c}_{01}}\phi_{11}(0) - i\Delta c\frac{\dot{\gamma}_1 - \dot{\gamma}_2}{\bar{c}_{01}^2}\phi_{01}(0) = \phi_{12}'(0) \qquad (51)$$

$$m\phi_{12}'''(0) - \phi_{11}'''(0) + ikR_1\{r[\bar{c}_{02}\phi_{02}'(0) + \dot{\gamma}_2\phi_{02}(0)] - [\bar{c}_{01}\phi_{01}'(0) + \dot{\gamma}_1\phi_{01}(0)]\} = ikR_1(1-r)G\phi_{01}(0)\bar{c}_{01}^{-1} \qquad (52)$$

The interfacial tension $\Gamma$ does not appear in the first-order approximation of the stream functions because the corresponding term is of order of $k^3$ (see Eq. (34)). Substituting Eq. (45) to the boundary conditions (47) – (52), we find the growth rate of the long-wave perturbations:

$$\Delta c = \frac{1}{3}kR_1 \frac{\bar{c}_{01}^2}{\dot{\gamma}_1} \frac{H_1 h_1(1) + H_2 h_1'(1) + H_3 h_2(-n) + H_4[h_2'(-n) - l_2 h_2''(-n)] + H_5 F}{4m^4 l_1^2 + m^3[2(3n+2)l_2 + m + 3n^2 + 4n]l_1 - 2n(m-1)[(3n+2)l_2 + (n^2+n)]} \quad . (53)$$

Here the following notations are introduced

$$H_1 = 3n\{4mn(3l_2 + n)l_1 + 2(3m + 6mn + 2n^2)l_2 + n(n^2 + 4mn + 3m)\} \qquad (53a)$$

$$H_2 = -3n\{4mn(3l_2+n)l_1 + 2(m+3mn+2n^2)l_2 + n(n^2+2mn+m)\} \tag{53b}$$

$$H_3 = 3r\left[4ml_1 + 2(3n+2)l_2 + 3n^2 + 4n + m\right] \tag{53c}$$

$$H_4 = 3rn\left[4ml_1 + n^2 + 2n + m\right] \tag{53d}$$

$$H_5 = n^2\{4m(3l_2+n)l_1 + (3m+4n)l_2 + n(m+n)\} \tag{53e}$$

It can be checked directly that in the case of the stick boundary conditions corresponding to $l_1 = l_2 = 0$ Eq. (53) coincides with the well-known result by Yih[24].

## V. DISCUSSION

### A. Propagation of the long interfacial waves

Consider at first the influence of the slip boundary conditions upon the details of propagation of the long-wave disturbances at the interface between the stratified fluids under shear. The general expression of the wave speed is represented by Eq. (42). It follows from this expression that in the absence of interfacial slip, $l_1 = 0$, the speed wave propagation coincides with the unperturbed flow velocity if both layers have the same viscosity (i.e. $m = 1$) even in the presence of slip at the bottom plate. On the contrary, the presence of interfacial slip would alter this rule. In the case of stick boundary conditions at the lower plate the following relative wave speed is obtained

$$\bar{c}_{01}(l_2 = 0) = \frac{m\{2n^2(m-1)(1+n) - 4m^2 l_1^2 - (m+4n+3n^2)ml_1\}}{(m+n+ml_1)\{4(m+n^3)ml_1 + m^2 + n^4 + 2mn(2+3n+2n^2)\}} \tag{54}$$

which takes a nonzero value at $m = 1$:

$$\bar{c}_{01}(l_2 = 0, m = 1) = -\frac{\{4l_1 + (1+4n+3n^2)\}l_1}{(1+n+l_1)\{4(1+n^3)l_1 + (1+n)^4\}} \tag{55}$$

This result is a consequence of the presence of non-degenerate interfaces even in a homogeneous fluid when slip exists between some of its parts.

Fig. 2 represents the dependence of the wave speed $c_0$ on the viscosity and the thickness ratios for two sets of boundary conditions: (a) sticky interface and slippery lower wall ($l_1 = 0$ and $l_2 = 1$) and (b) slippery interface and the sticky lower wall ($l_1 = 1$ and $l_2 = 0$). The dashed lines correspond to the stick boundary conditions with $l_1 = l_2 = 0$.[24] It is seen from Fig. 2a that the slip at the lower wall results in an acceleration of the propagation of disturbances in comparison with the sticky boundary conditions. This difference is rapidly growing with decreasing the thickness ratio. In the limiting case of a very thin lower layer, $n \sim 0$, the wave speed is of the order of the basic fluid velocity, $c_0 \sim V_1 = l_2 \dot{\gamma}_2$. In contrast to the wall slip, the interfacial slip leads just to a small altering of the wave speed. Nevertheless, these changes are somewhat complicated: Fig. 2b shows that the wave speed could exceed one for the stick boundary conditions and for small viscosity ratios while it becomes smaller if $m$ is sufficiently large. The transition between these two regimes depends on the thickness ratio: the larger the value of $n$, the larger the value of $m$ corresponding to the threshold between these states.

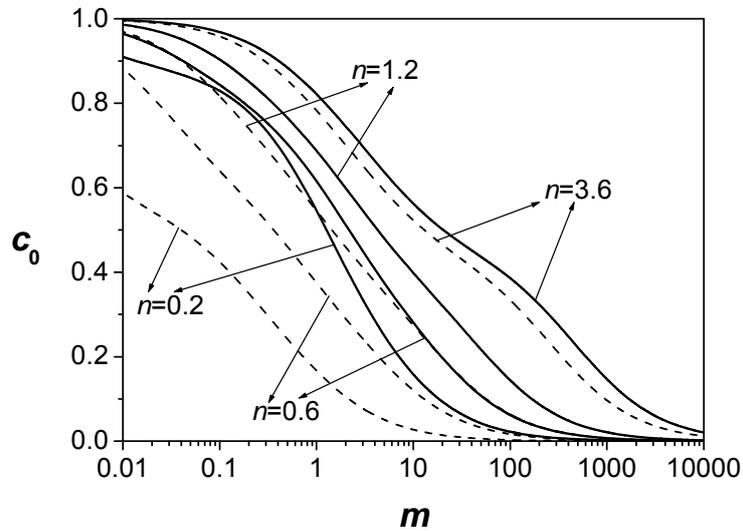

(a)

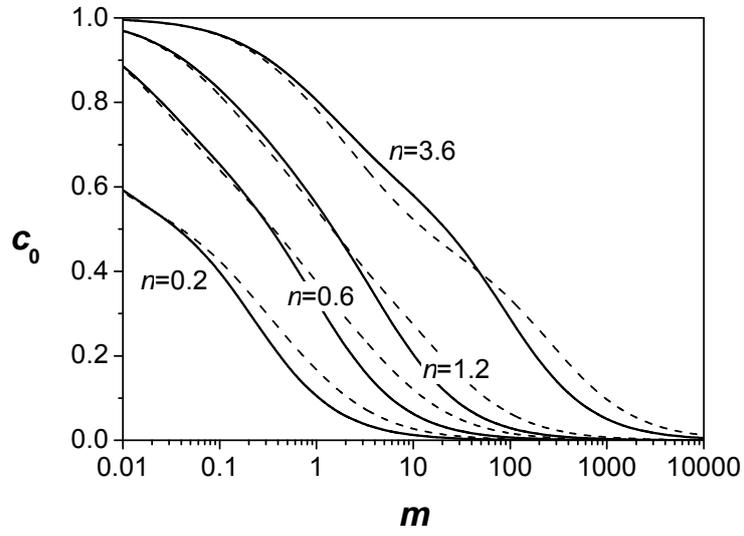

**(b)**

**Fig. 2.** The dependence of velocity of interfacial wave propagation $c_0$ on the viscosity ratio $m$ at different thickness ratios $n$. Solid lines: (a) $l_1 = 0$, $l_2 = 1$ and (b) $l_1 = 1$, $l_2 = 0$. The dashed lines correspond to the non-slip boundary conditions ($l_1 = l_2 = 0$).[24]

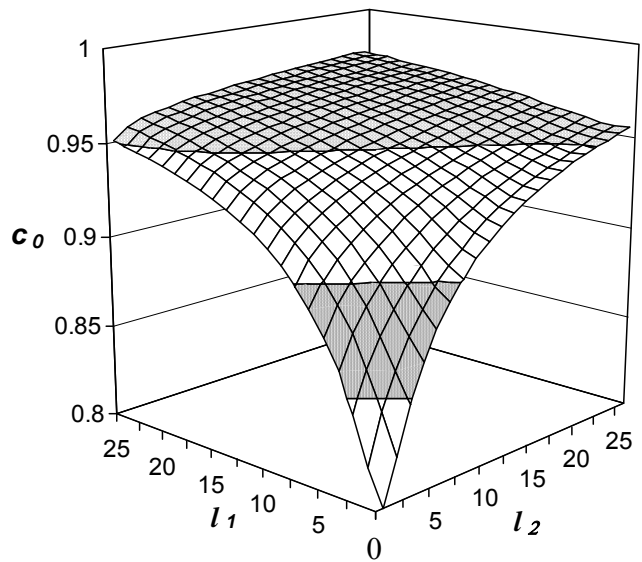

**Fig. 3.** The dependence of velocity of interfacial wave propagation $c_0$ on the slippage lengths $l_1$ and $l_2$. The viscosity and the thickness ratios are equal to $m = 1.5$ and $n = 5$.

Fig. 3 shows an example of the combined effect of the interfacial and the wall slip on the propagation of the disturbances at $m = 1.5$ and $n = 5$. We can see that the increase in the slippage lengths $l_1$ and $l_2$ results in the monotonous growth of the wave speed with a limit equal to the velocity of the upper plate. The same asymptotic limit is found for the primary interface velocities $V_1$ and $V_2$ (cf. Eqs. (13), (14)) which are obviously correlated with the speed of the long-wave perturbations.

**B. Stability analysis.**

Now we discuss peculiarities of the hydrodynamic stability of the sheared two-layer system caused by slip boundary conditions. The stability is characterized by the perturbation growth rate $\Delta c$ given by Eq. (53): positive values of $\Delta c$ promote growth of wave amplitude thus leading to unstable flow while negative values of $\Delta c$ reduce the interfacial disturbances resulting in stable flow. The neutral stability corresponds to $\Delta c = 0$.

We begin with the density-matched fluids corresponding to $r = 1$. Fig. 4 shows the behavior of the growth rate as a function of the viscosity ratio $m$ corresponding to the slippery lower wall, $l_2 = 1$, and the non-slip interface between the layers, $l_1 = 0$, in comparison with that for sticky boundary conditions. Two ranges of thickness ratio are considered in these plots: (a) $n \leq 1$ (the thickness of the upper layer is larger or equal to the lower one) and (b) $n > 1$ (the lower layer is thicker than the upper one). We can see that the deviation of the growth rate from that at the sticky boundary conditions[24] depends significantly on the values of viscosity and thickness ratios. For instance, at $n \leq 1$ the growth rate increases when the upper layer is more viscous than the lower one, i.e. in the range of $m < 1$: the lower the thickness ratio, the higher the values of $\Delta c$. In other words, the wall slip could induce the interfacial instability at $n \leq 1$ and $m < 1$ by shifting the perturbation growth rate towards the positive domain. If the lower fluid is denser than the upper one, $m > 1$, wall slip results in shifting the growth rate curves towards larger values of the viscosity. This deviation increases with the increase of thickness ratio (see Fig. 4a and 4b). Subsequent increase in the viscosity ratio leads to neutral stability at any value of $n$.

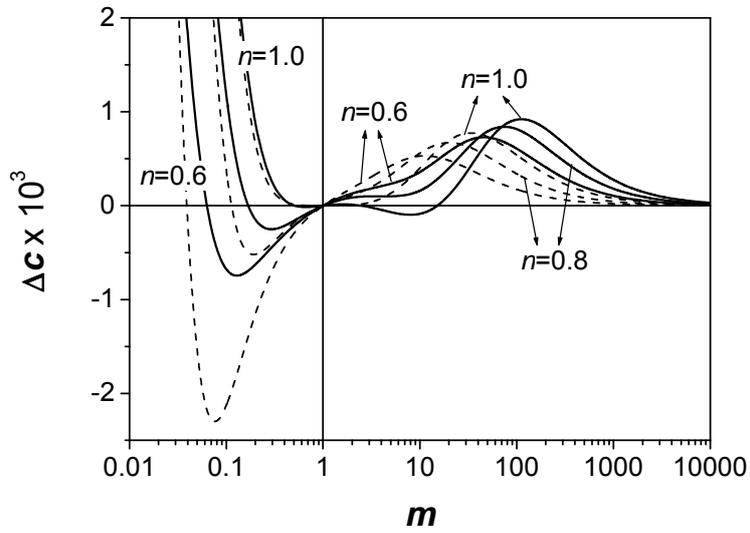

(a)

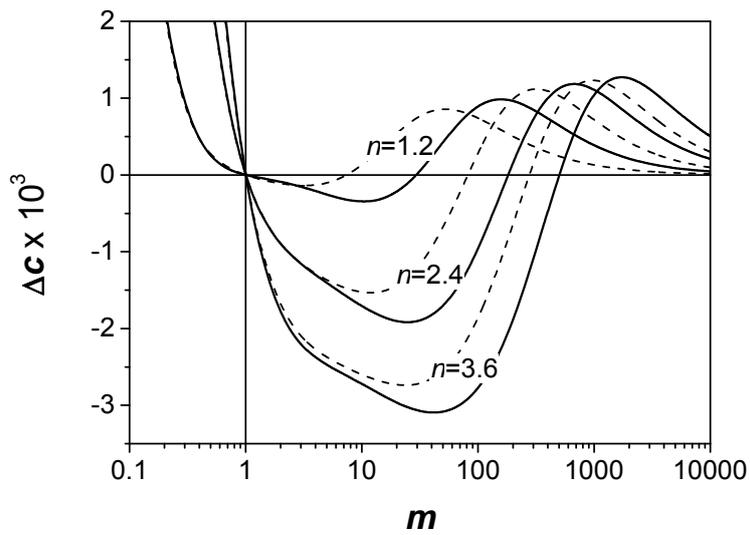

(b)

**Fig. 4.** The dependence of the growth rate $\Delta c$ of interfacial perturbations on the viscosity ratio $m$ under the wall slip ($l_2 = 1$) and the stick interface boundary conditions. Solid curves: (a) $n \leq 1$ and (b) $n > 1$. The dashed curves correspond to the stick boundary conditions ($l_1 = l_2 = 0$).[24] The density ratio is $r = 1$.

The effect of slip at the interface between the viscous layers in the absence of wall slip is presented in Fig. 5 for $l_1 = 0.1$. In contrast to the previous case, the interfacial slip leads to increase the growth rate in the whole range of viscosity and thickness ratios. The transition from stable to unstable flow can be observed in the domain of $n < 1$ and $m < 1$ (Fig. 5a) and at $n > 1$ and $m > 1$ (Fig. 5b) demonstrating the importance of interfacial slip. But the most exciting outcome of Fig. 5 is that the interfacial slip can lead to the hydrodynamic instability of the stratified system with matched viscosities and densities of the layers. In this case Eq. (53) is transformed to

$$\Delta c = \frac{1}{60} k R_1 n^2 l_1 \{1536 n (n^4 - n^3 + n^2 - n + 1) l_1^4 - 16(1+n)(3n^6 - 142n^5 + 65n^4 - 135n^3 + 65n^2 - 142n + 3) l_1^3 - 12(1+n)^2 (2n^6 - 94n^5 - 57n^4 - 94n^3 - 57n^2 - 94n + 2) l_1^2 - 3(1+n)^5 (n^4 - 78n^3 - 10n^2 - 78n + 1) l_1 + 16n(1+n)^8 \} \Big/ \{(1+n)^2 [4(n^2 - n + 1) l_1 + (1+n)^3]^3 (1+n+l_1)^2\} \quad (56)$$

However, the shear flow will be stable if $n \ll 1$. As a matter of fact, an expansion of Eq. (56) over small $n$ values gives the following expression

$$\Delta c = -\frac{1}{20} \frac{k R_1 l_1^2}{(4l_1 + 1)(l_1 + 1)^2} n^2 + O(n^3) \quad (57)$$

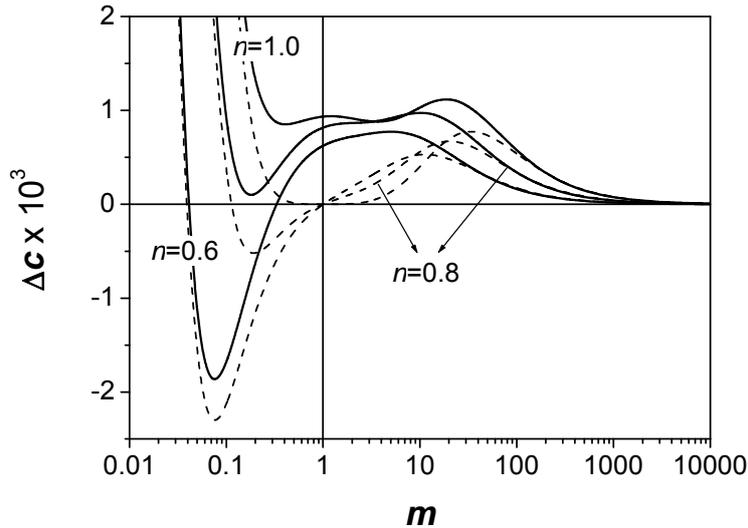

(a)

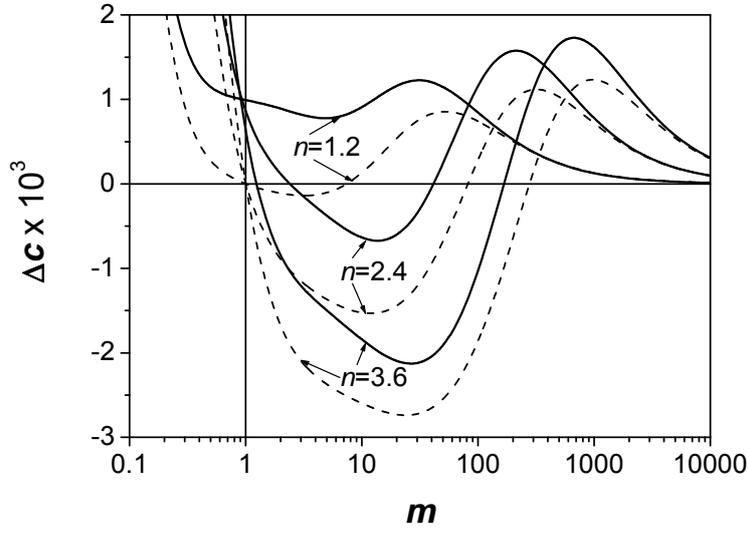

**(b)**

**Fig. 5.** The dependence of the growth rate $\Delta c$ of interfacial perturbations on the viscosity ratio $m$ with the interfacial slip ($l_1 = 0.1$) and the stick wall boundary conditions. Solid curves: (a) $n \leq 1$ and (b) $n > 1$. The dashed curves correspond to the stick boundary conditions ($l_1 = l_2 = 0$).[24] The density ratio is $r = 1$.

which is negative. Thus, we can conclude that the presence of slip between two parts of a homogeneous viscous fluid would result in unstable shear flow if the thickness ratio of these two parts is not too small. On the other hand, the flow is stable if the lower layer is much thinner than the upper one.

The combined effect of slip boundary conditions at the interface and at the lower wall is demonstrated in Fig. 6 showing the growth rate as a function of the two slippage lengths $l_1$ and $l_2$. The thickness and viscosity of the lower layer are taken larger than those of the upper layer ($m = 1.5$ and $n = 5$). These ratios guarantee a stable flow under the sticky boundary conditions. It is seen that increasing $l_1$ leads to a sharp increase of the growth rate at small values of the wall slippage length $l_2$. As a result, the initially negative sign of $\Delta c$ could become positive at a definite value of the interfacial slippage length $l_1$, thus leading to unstable flow. On the other hand, the influence of the interfacial slip is reduced by the wall slip: the slope of the growth rate with respect to $l_1$ - decreases with increasing $l_2$. Nevertheless, the transition from stable to unstable flow is observed in the whole range of the wall slippage length if $l_1$ exceeds a threshold depending on $l_2$. The increase in $l_2$ at fixed value of $l_1$ results in a reduction of the perturbation growth rate towards neutral stability, $\Delta c = 0$, of the chosen system. This behavior is peculiar to both the positive and negative domains of the growth rate. On the contrary, the increase in the wall slippage length would promote the interfacial instability if the lower layer is less viscous than the upper one (c.f. Fig. 4).

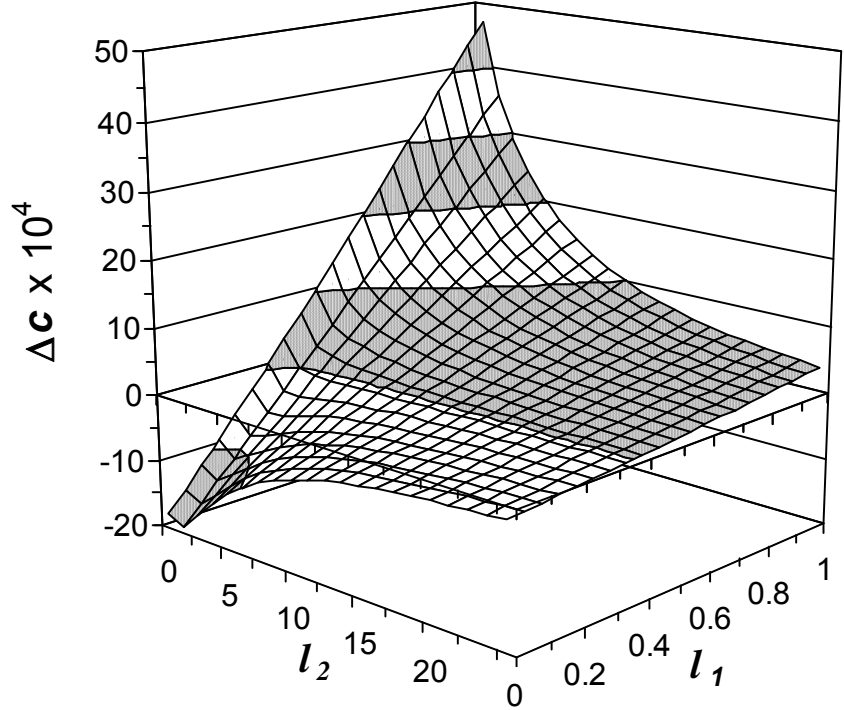

**Fig. 6.** The dependence of the growth rate $\Delta c$ of the interfacial perturbations for the density-matched fluids ($r = 1$) on the slippage lengths $l_1$ and $l_2$. The viscosity and the thickness ratios are equal to $m = 1.5$ and $n = 5$.

To complete the picture, consider the perturbation growth rate as a function of the slippage lengths for small viscosity and thickness ratios, $m = 0.05$ and $n = 0.6$. For the non-slip boundary conditions these values correspond also to a stable interface. Fig. 7 shows that increasing the wall slippage length may result in non-monotonic behavior of $\Delta c$ as a function of $l_2$. This is inherent to small values of the thickness ratio whereby wall slip promotes an increase of the growth rate (see Fig. 4a). Further enhancement of $l_2$ shifts the growth rate to the positive domain and then drives it towards neutral stability (cf. Fig. 6).

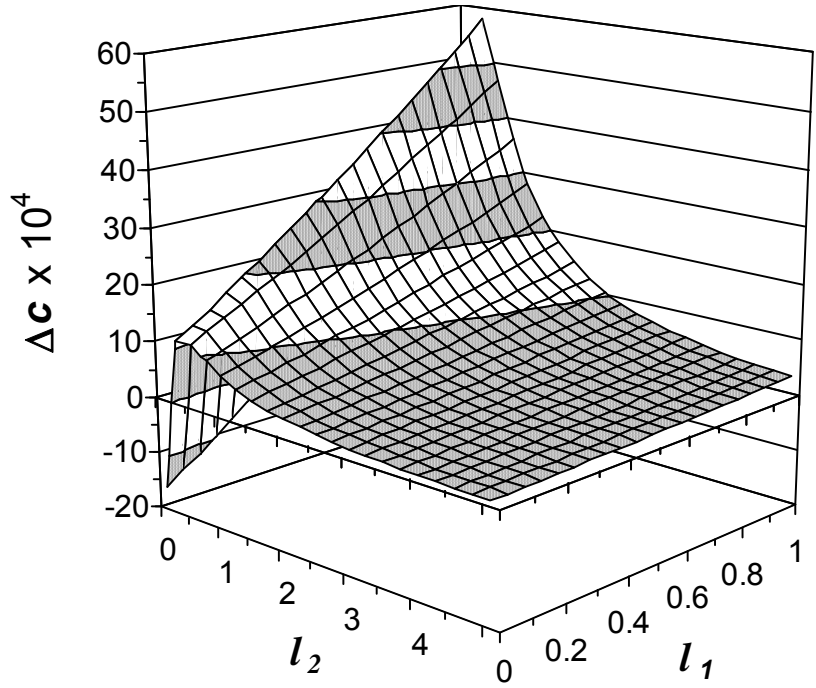

**Fig. 7.** The same as in Fig. 6 but viscosity and thickness ratios are equal to $m = 0.05$ and $n = 0.6$.

The influence of gravity on the hydrodynamic stability of the two-layer fluid with slip boundary conditions is represented in Fig. 8 (the viscosity and thickness ratios are the same as in Fig. 6). The first example shown in Fig. 8a concerns the system where the lower layer has a higher density than the upper one, i.e. $r = 2$. One can see that the domain of stable flow increases in comparison with that of the density-matched system (cf. Fig. 6). Nevertheless, even in this case, the interfacial slip could induce unstable flow by shifting $\Delta c$ to the positive area. If the upper layer has a higher density than the lower one, $r < 1$, the interface is always unstable. This is shown in Fig. 8b for $r = 0.5$. The obtained solutions are consistent with the Rayleigh-Taylor theory describing the stability of inviscid fluids. Increasing the lower wall slippage length drives the perturbation growth rate towards the constant limit in both cases. The only difference is that the asymptotic value of $\Delta c$ is negative at $r > 1$ and positive at $r < 1$.

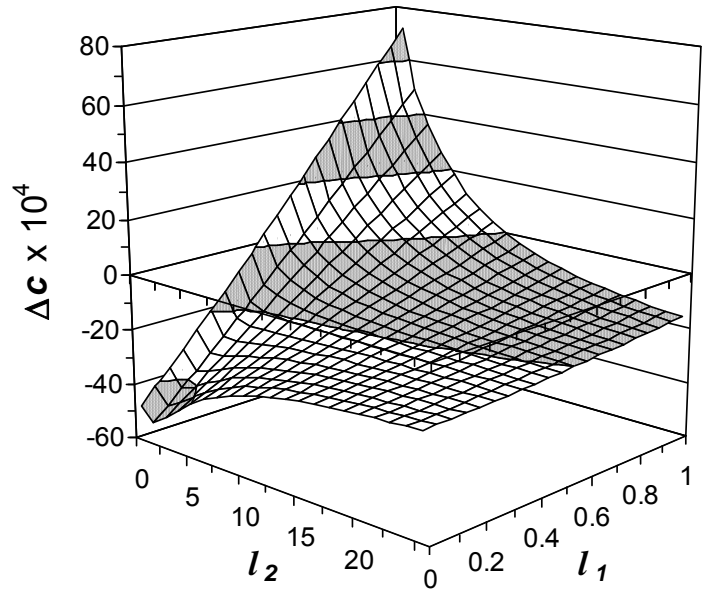

(a)

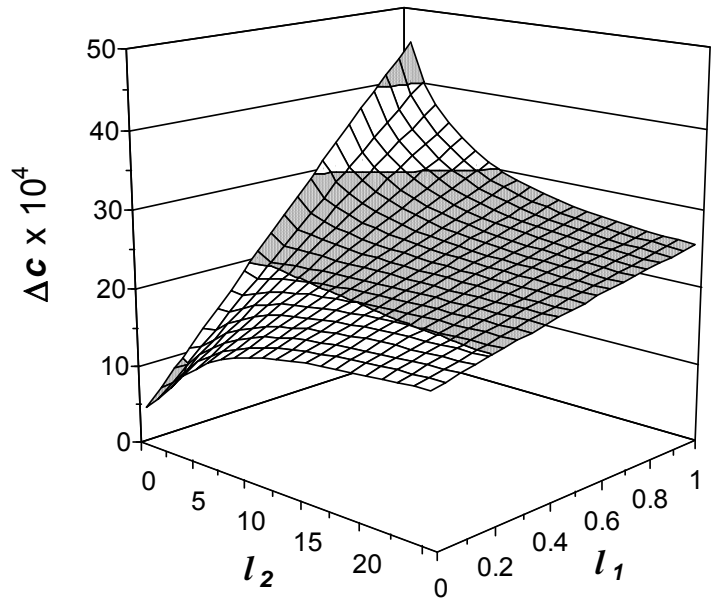

(b)

**Fig. 8.** The same as in Fig. 6 but for different density ratios: (a) $r = 2$ and (b) $r = 0.5$.

## V. CONCLUDING REMARKS

The influence of slip boundary conditions on the long-wave hydrodynamic stability of a two-layer system of immiscible and incompressible viscous layers subjected to simple shear flow has been studied. It was found that slip at the interface between the layers and the outer wall could considerably alter the flow stability. The regimes of flow are shown to be defined by the slippage lengths, which characterize the discontinuity of interfacial and wall velocities, along with viscosity, thickness, and density ratios of the layers.

The slip between the layers results in an increase of the growth rate of the long-wave perturbations in all range of the viscosity and thickness ratios thus making wider the area of unstable flow. This effect can lead to the hydrodynamic instability even for a homogeneous fluid containing two slipping parts, if their thickness ratio is not too small. Practically this may be realized if a slippery flexible membrane was immersed into the fluid. However, the shear flow is predicted to be stable if the lower layer is thin enough.

The slip at the lower wall increases the ranges of viscosity and thickness ratios corresponding to the stable and unstable flows for the system with the less viscous upper layer. The following increase of the wall slippage length leads to neutral stability in the case of density-matched fluids. However, if the lower layer is of larger (smaller) density than the upper one, the wall slip leads to stable (unstable) shear flow. If the lower layer is of smaller viscosity than the upper one, the wall slip promotes the interfacial instability.

The slip boundary conditions affect the propagation of interfacial perturbations. Wall slip results in the increase of the speed of long waves in the whole range of viscosity ratio. On the contrary, interfacial slip tends to decrease (increase) the wave speed when the lower layer is sufficiently more (less) viscous than the upper one. Increase of the interfacial and wall slips makes the wave speed approach the velocity of the upper plate.

**ACKNOWLEGMENTS**

This work was supported by the Russian Foundation for Basic Researches under the award number 05-03-33018-a.

**APPENDIX**

The coefficients of the zero order stream functions could be found from the boundary conditions (29), (30), (38) – (41) by substituting of Eq. (37). Putting $A_1 = 1$, we obtain the following set of equations:

$$\left. \begin{array}{c} 1 + B_1 + C_1 + D_1 = 0, \; B_1 + 2C_1 + 2D_1 = 0, \\[6pt] A_2 - nB_2 + n^2 C_2 - n^3 D_2 = 0, \\[6pt] B_2 - 2(n + l_2)C_2 + 3(n^2 + 2nl_2)D_2 = 0, \\[6pt] \bar{c}_{01} A_2 = \bar{c}_{02}, \; C_1 - mC_2 = 0, \; D_1 - mD_2 = 0, \\[6pt] B_1 - B_2 - 2l_1 C_1 + \dfrac{\dot{\gamma}_1 - \dot{\gamma}_2}{\bar{c}_{01}} = 0 \end{array} \right\} \qquad (A1)$$

The solutions of Eq. (A1) are

$$A_2 = 1 + m l_1 \dot{\gamma}_2 \bar{c}_{01}^{-1} \qquad (A2)$$

$$B_1 = -E^{-1} \left\{ 6m^2 l_1^2 + 6m\left[(2n+1)l_2 + n + n^2\right]l_1 - (m-1)\left[6n(2n+1)l_2 + \left(m + 3n^2 + 4n^3\right)\right] \right\} \qquad (A3)$$

$$B_2 = B_1 - 2l_1 m C_2 + (m-1)\dot{\gamma}_2 \bar{c}_{01}^{-1} \qquad (A4)$$

$$C_1 = mC_2 \qquad (A5)$$

$$C_2 = (mE)^{-1} \left\{ 3m\left[2nl_2 - m + n^2\right]l_1 - 2(m-1)\left[3n^2 l_2 + m + n^3\right] \right\} \qquad (A6)$$

$$D_1 = mD_2 \qquad (A7)$$

$$D_2 = (mE)^{-1} \left\{ 2m^2 l_1^2 + 2m(l_2 + m + n)l_1 - (m-1)\left(2nl_2 - m + n^2\right) \right\} \qquad (A8)$$

The coefficient *E* is equal to

$$E = 4m^2 l_1^2 + m\left[2(3n+2)l_2 + m + 4n + 3n^2\right]l_1 - 2(m-1)n\left[(3n+2)l_2 + n + n^2\right] \tag{A9}$$

The relative speed of the long-wave propagation $\bar{c}_{01}$ follows from Eq. (A1) and is presented in Eq. (42).

**References**

[1] Migler, K.B., H. Hervet and L. Leger, "Slip transition of a polymer melt under shear stress," Phys. Rev. Lett. **70**, 287 (1993).

[2] Archer, L.A., R.G. Larson and Y.-L Chen., "Direct measurements of slip in sheared polymer solutions," J. Fluid Mech. **301**, 133 (1995).

[3] Hatzikiriakos, S. G., I. B. Kazatchkov, and D. Vlassopoulos, "Interfacial phenomena in the capillary extrusion of metallocene polyethylenes", J. Rheol. **41**, 1299 (1997).

[4] Wise, G.M., M.M. Denn and A.T. Bell, "Surface mobility and slip of polybutadiene melts in shear flow," J. Rheol. **44**, 549 (2000).

[5] Wang, S.Q., "Molecular transitions and dynamics of polymer/wall interfaces: Origins of flow instabilities and wall slip," Adv. Polym. Sci. **138**, 227 (1999).

[6] Denn, M.M., "Extrusion instabilities and wall slip," Ann. Rev. Fluid. Mech. **33**, 265 (2001).

[7] Robert, L., B. Vergnes and Y. Demay, "Flow birefringence study of the stick-slip instability during extrusion of high-density polyethylenes," J. Non-Newt. Fluid Mech. **112**, 27 (2003).

[8] Zhao, R. and C. W. Macosko, "Slip at polymer-polymer interfaces: Rheological measurements on coextruded multilayers," J. Rheol. **46**, 145 (2002).

[9] Churaev, N.V., V.D. Sobolev and A.N. Somov, "Slippage of liquids over lyophobic solid surfaces," J. Colloid Interface Sci. **97**, 574 (1984).

[10] Bocquet, L. and J.L. Barrat, "Hydrodynamic boundary conditions and correlation functions of confined fluids," Phys. Rev. Lett. **70**, 2726 (1993)

[11] Vinogradova, O.I., "Slippage of water over hydrophobic surfaces," Int. J. Miner. Process. **56** 31 (1999).

[12] Lin, C.C., "A mathematical model for viscosity n capillary extrusion of two components polyblends," Polym. J. (Tokyo) **11**, 185 (1979).

[13] Miroshnikov, Y.P. and E.N. Andreeva, "Flow of polypropylen-polysterene blends having coaxial phase structure," Vysokomol. Soedin., Ser. A 29, 579 (1987).

[14] Bousmina, M.J., J.F. Palierne and L.A. Utracki, "Modeling of structured polyblend flow in a laminar shear field," Polym. Eng. Sci. **39**, 1049 (1999).

[15] De Gennes, P.G., "Viscometric flow of tangled polymers," C. R. Acad. Sci. Paris, Ser. B **288**, 219 (1979).

[16] Brochard, F. and P.G. de Gennes, "Shear dependent slippage at a polymer/solid interface," Langmuir **8**, 3033 (1992).

[17] Brochard, F., P.G. de Gennes and S. Troian, "Slippage at the interface between two slightly incompatible polymers," C. R. Acad. Sci. Paris, Ser. II: Mech., Phys., Chim., Sci. Terre Univers **310**, 1169 (1990).

[18] De Gennes, P.G., "Mechanical properties of polymer interfaces," in Physics of Polymer Surfaces and Interfaces, edited by I.C. Sanchez (Butterworth-Heinemann, MA, 1992), pp. 55.

[19] Goveas, J.L. and G.H. Fredrickson, "Apparent slip at a polymer-polymer interface," Eur. Phys. J. B **2**, 79 (1998).

[20] Kalika, D.S. and M.M. Denn, "Wall lip and extrudate distortion in linear low-density polyethylene," J. Rheol. **31**, 815 (1987).

[21] Ajji A., H.P. Schreiber and D. Duchesne, "Flow defects in linear low density polyethylene processing: Instrumental detection and molecular weight dependence," Polym. Eng. Sci. **33** 1524 (1993).

[22] Black, W.B. and M.D. Graham, "Wall-slip and polymer-melt flow instability," Phys. Rev. Lett. **77,** 956 (1996).

[23] Black, W.B. and M.D. Graham, "Effect of wall slip on the stability of viscoelastic plane shear flow," Phys. Fluids **11**, 1749 (1999).

[24] Yih, C.-S., "Instability due to viscosity stratification," J. Fluid Mech. **27**, part 2, 337 (1967).

[25] Renardy, Y., "Stability of the interface in two-layer Couette flow of upper convected Maxwell liquids," J. Non-Newt. Fluid Mech. **28**, 99 (1988).

[26] Yiantsios, S. G. and B. G. Higgins, "Linear stability of plane Poiseuille flow of two superposed fluids," Phys. Fluids **31**, 3225 (1988).

[27] Hooper, A.P., "The stability of two superposed viscous fluids in a channel," Phys. Fluids A **1**, 1133 (1989).

[28] Khomami, B., "Interfacial stability and deformation of two stratified power law fluids in plane Poiseuille flow, Part I. Stability analysis," J. Non-Newt. Fluid Mech. **36**, 289 (1990).

[29] Charru, F. and E.J. Hinch, "'Phase diagram' of interfacial instabilities in a two-layer Couette flow and mechanism of the long-wave instability," J. Fluid Mech. **414**, 195 (2000).